\documentclass[12pt]{iopart}

\expandafter\let\csname equation*\endcsname\relax
\expandafter\let\csname endequation*\endcsname\relax
\usepackage{amsmath}
\usepackage{xspace}
\usepackage{amssymb}
\usepackage{tabularx}
\usepackage{graphicx}
\usepackage{url}         
\usepackage{lscape}      
\usepackage{multicol}    
\usepackage{cancel}      
\usepackage{lettrine}
\usepackage{float}
\usepackage{pdfpages}
\usepackage{booktabs}
\usepackage{bm}
\usepackage{dcolumn}
\usepackage{etoolbox}
\usepackage{epstopdf, epsfig}

\begin{document}

\title[Transition to chaos in rotating RBC]{Transition to chaos and magnetic field generation in rotating Rayleigh-Bénard convection}

\author{Dalton N. Oliveira}
\address{Aeronautics Institute of Technology – ITA, 12228-900, S\~ao Jos\'e dos Campos, SP, Brazil}
\ead{dalton@ita.br}
\author{Roman Chertovskih}
\address{Research Center for Systems and Technologies (SYSTEC), ARISE,  Faculdade de Engenharia da Universidade do Porto, Rua Dr. Roberto Frias, s/n 4200-465, Porto, Portugal }
\ead{roman@fe.up.pt}
\author{Erico L. Rempel}
\address{Aeronautics Institute of Technology – ITA, 12228-900, S\~ao Jos\'e dos Campos, SP, Brazil}
\ead{rempel@ita.br}
\author{Francis F. Franco}
\address{Federal University of Jataí - UFJ, 75801-615, Jataí, GO, Brazil\\
}
\ead{francis\_franco@ufj.edu.br}

\date{\today}

\begin{abstract}
Hydrodynamic and magnetohydrodynamic convective attractors in three-dimensional rotating Rayleigh-Bénard convection 
are studied numerically by varying the Taylor and Rayleigh numbers as control parameters. First, an analysis of hydrodynamic attractors and their bifurcations is conducted, 
where routes to chaos via quasiperiodicity are identified. Second, the behaviour of the magnetohydrodynamic system is investigated by introducing a seed magnetic field and measuring its growth or decay as a function of the Taylor number, while keeping the Rayleigh number fixed. Analysis of the attractors  shows that rotation has a significant impact on magnetic field generation in Rayleigh-Bénard convection, with the critical magnetic Prandtl number changing nonmonotonically with the rotation rate. It is argued that a nonhysteretic blowout bifurcation with on-off intermittency is responsible for the transitions to dynamo.
\end{abstract}

\maketitle

\section{Introduction}
\label{introduction}

Thermal convection in a rotating fluid is a physical phenomenon of interest for geophysics and astrophysics, playing a key role in magnetic field generation in planets and stars. In the geodynamo, the magnetic field is generated by the thermal convection process in the Earth's outer liquid iron core~\cite{KRAUSE1980}. The conducting fluid dynamics is strongly influenced by a combination of the Coriolis force, which is due to the Earth's rotation, and the Lorentz force, describing the influence of the magnetic field on the flow~\cite{Verhoogen1961heat, Glatzmaier1998dynamo}. In the Sun, turbulent flows are also strongly influenced by the Coriolis and Lorentz forces involved in the magnetohydrodynamic processes in the convective zone~\cite{PRIEST2014}. 
Dynamo theory provides a framework describing the origin and morphology of these magnetic fields and their spatial and temporal dynamics ~\cite{parker2019cosmical}. Dynamo theory investigates the hydromagnetic processes by which an electrically conducting fluid 
may amplify a weak magnetic field and sustain it; in the presence of differential rotation, this process gives rise to the so-called $\alpha-\Omega$ dynamo~\cite{Brandenburg2018}. However, dynamos without differential rotation can also generate magnetic fields, where the dynamo works solely based on turbulent motions, thus generating the $\alpha^{2}$ type dynamo~\cite{rincon2019dynamo}.

Turbulent rotating convection is believed to be the primary mechanism for the generation of large-scale magnetic fields observed in stars and planets according to dynamo theory. The Earth's rotational velocity is relatively high; the geodynamo in the outer core is assumed to operate in the magnetostrophic regime, i.e., when the Lorentz force is of the order of the Coriolis force \cite{merrill1998magnetic}. Rotation is also an important factor in the physics of the solar tachocline, a transition layer between the radiative and convective zones where the differential rotation rate varies rapidly and where it is assumed that the Sun's large-scale magnetic field is generated ~\cite{Hughes2007solar}. The differential rotation leads to the formation of $\Omega$-effect dynamos ~\cite{MOFFAT1978}. Although slow in contrast to the fast dynamos thought to operate in stars, this mechanism is considered a key element in the theory of the solar dynamo ~\cite{rincon2019dynamo}. Boundary layers and shear flows that develop in rapidly rotating fluids are the structures that control the dynamics of fluid processes and planetary dynamos ~\cite{busse2007dynamics}. Therefore, the question of how the magnetic field generation processes are affected by the rotational velocity is of interest for geophysical and astrophysical applications.

A simple setup to study thermal convection is the rotating Rayleigh-Bénard convection (RBC), which consists of a rotating layer of fluid confined between two horizontal planes, heated from below and cooled from above.
Despite its simplicity, realistic flows, such as convective rolls, are observed in RBC. Such flows are present in systems of geophysical and astrophysical nature~\cite{glatzmaier2013introduction, yan2021scaling, hsia2022route, kumar2022anisotropy, samuel2022large}. 
Nonlinear dynamics and chaos theory constitute an important tool in understanding several physical phenomena occurring in RBC models. In the absence of rotation, Chertovskih et al.~\cite{Chertovskih2015route} studied the transition to hyperchaotic regimes, going through quasiperiodic regimes, coexistence of attractors, and intermittent chaotic attractors. Chimanski et al. \cite{Chimanski2016off} further explored 
 this route to hyperchaos and the presence of chaotic saddles. 
 
The dependence of magnetic field generation on the rotation rate has been explored in previous studies of rotating convection. 
 Meneguzzi and Pouquet 
 \cite{meneguzzi1989turbulent} investigated magnetic field generation in the absence and presence of rotation, considering fully developed turbulent convection. They observed that rotation can be beneficial for nonlinear dynamos, as the critical magnetic Reynolds number decreased when rotation was active. However, Cattaneo and Hughes \cite{cattaneo2006dynamo} suggested that rotation is not a significant factor for magnetic field generation, with similar growth rates and saturation levels being found in systems with and without rotation. 
 Although convective turbulent nonlinear dynamos do not require rotation, near the onset of convection, rotation seems to be a necessary element for dynamo action ~\cite{matthews1999dynamo}. Chertovskih et al. ~\cite{Chertovskih2010} found a more complex dependence of magnetic field generation on the rotation rate. Near the onset of convection, in the nonlinear dynamo regime, magnetic energy was shown to depend on the Taylor number in a nonmonotonously manner.
In nonrotating RBC, Chertovskih et al. \cite{Chertovskih2017} showed that quasiperiodic hydrodynamic flows are more beneficial for dynamo action than chaotic flows. 
 The asymptotic scaling behavior of convection-driven dynamos in rapidly rotating RBC was studied as a function of several parameters by Calkins et al. \cite{calkins22}, including transition from large-scale to small-scale dynamos as a function of the magnetic Reynolds number. Although RBC does not provide a realistic setup for the simulation of turbulent convection in a geophysical or astrophysical context, it continues to be employed in the investigation of basic phenomena present in natural dynamos due to its simplicity. 

In general, high Rayleigh numbers favor magnetic field generation, as noted in spherical shell simulations \cite{mather2021regimes}, i.e., the critical magnetic Prandtl number $(P_{m})$ for dynamo action decreases with increasing Rayleigh number $(R)$. For moderate values of $R$, a similar dependence was found for plane layer dynamos, as reported in \cite{yan2022asymptotic}. The motivation of the present study is to investigate magnetic field generation in RBC with an electrically conductive fluid in the presence of rigid rotation. For this, we study hydrodynamic regimes as a function of the Rayleigh number ($R$), 
 indicating the magnitude of thermal buoyancy forces, and the Taylor number ($Ta$), measuring the speed of rotation. A series of bifurcation diagrams are constructed and transitions to hydrodynamic chaos are identified. We then investigate magnetic field generation in this RBC model 
 and identify intermittent hydromagnetic chaotic attractors similar to those reported by Sweet et al. \cite{Sweet2001blowout}, Spiegel  \cite{spiegel2009chaos}, and Rempel et al. \cite{rempel2009novel}, found in MHD simulations in a periodic box with helical ABC-forcing without rotation. Sweet et al. \cite{sweet2001alowout, Sweet2001blowout} demonstrated that in this setup transition to dynamo takes place as a result of a nonhysteretic blowout bifurcation. Likewise, here a nonhysteretic blowout bifurcation is suggested as the mechanism for transition to dynamo in rotating RBC. The article is organized as follows: In section \ref{section-2} we present the equations governing the MHD system, the boundary conditions, and the numerical methods used. In section \ref{section-3}, we present the results obtained from hydrodynamic simulations for  various values of $R$ and $Ta$. In section \ref{section-4}, we consider the MHD system by adding a seed magnetic field and study how rotation influences transition to dynamo; the conclusions are presented in section \ref{section-5}.

\section{GOVERNING EQUATIONS AND NUMERICAL METHODS} \label{section-2}

We adopt the simulation model employed in references ~\cite{Chertovskih2010, Chertovskih2017}. An incompressible fluid heated from below in a plane horizontal layer rotating about the vertical axis is considered. In a Cartesian reference frame with the orthonormal basis $\left ( \mathbf{e}_x, \mathbf{e}_y, \mathbf{e}_z \right )$, where  $\mathbf{e}_z$ is opposite to the direction of gravity, the  equations governing  the magnetohydrodynamic system  are:
\begin{gather}
 \frac{\partial \textbf{v}}{\partial t} = \textbf{v} \times \left ( \nabla \times \textbf{v} \right ) + P \nabla^{2} \textbf{v} + PR \theta \mathbf{e}_{z} + P\sqrt{Ta}\hspace{0.075cm} \textbf{v} \times \mathbf{e}_{z} - \nabla p- \mathbf{B} \times \left ( \nabla \times \mathbf{B} \right ), \label{eqn:1}  \\ 
\frac{\partial \mathbf{B}}{\partial t}  = \nabla \times \left ( \textbf{v} \times \mathbf{B} \right ) + \frac{P}{P_{m}}\nabla^{2}\mathbf{B}, \label{eqn:2} \\
\frac{\partial \theta}{\partial t}  = \nabla^{2} \theta - \left (\textbf{v} \cdot \nabla \right ) \theta + \mathit{v_{z}}, \label{eqn:3}\\
\nabla \cdot \textbf{v} = 0, \qquad    \nabla \cdot \mathbf{B} = 0. \label{eqn:4}
\end{gather}
where $\textbf{v}(\mathbf{x},t) = ( v_{x}, v_{y}, v_{z})$ is the velocity field, $\mathbf{B}(\mathbf{x},t) = ( b_{x}, b_{y}, b_{z})$ is the magnetic field, $p(\mathbf{x}, t)$ is the pressure, and $\theta (\mathbf{x}, t) = T(\mathbf{x}, t) - (T_{1} + (T_{2} - T_{1}) z)$  is the difference between the temperature of the fluid and the linear temperature profile.  
The governing equations of the RBC in the presence of magnetic field are characterized in the dimensionless form by the Prandtl number $P$ (defined as the ratio between the kinematic viscosity of the fluid and the thermal diffusivity), the magnetic Prandtl number $P_{m}$ (defined as the ratio between the viscosity of the fluid and the magnetic diffusivity coefficient), the Rayleigh number $R$ (representing the magnitude of the thermal buoyancy forces), and the Taylor number $Ta$ (proportional to the rotational velocity). They are defined as
\begin{equation}
P = \frac{\nu}{\kappa}, \quad  P_{m} = \frac{\nu}{\eta}, \quad R = \frac{\alpha \textsl{g}\delta T d^{3}}{\nu \kappa}, \quad Ta = \frac{4 \Omega^2 d^4}{\nu^2}, \label{eqn:5}
\end{equation}
where $\nu$ is the kinematic viscosity, $\kappa$ is the thermal diffusivity, $\eta$ is the magnetic diffusivity, $\alpha$ is the thermal expansion coefficient, $\textsl{g}$ is the acceleration due to gravity, $\Omega$ is the angular velocity, $\delta T$ is the temperature difference between the layer boundaries, and $\emph{d}$ is the vertical size of the layer. The units for length and time  are $\emph{d}$ and the vertical heat diffusion time, $\emph{d}^{\hspace{0.075cm} 2} / \kappa$; $\textit{\textbf{v}}, \mathbf{B}$ and $\theta$ are measured in units of $\kappa / \emph{d}$, $\sqrt{\mu_{0}\rho}\kappa / \emph{d}$ and $\delta T$, respectively. Here, $\mu_{0}$ stands for the magnetic permeability in a vacuum and $\rho$ for the mass density.  These dimensional parameters can differ significantly (by many orders of magnitude) in geophysical and astrophysical fluids. For example, the Prandtl number in the solar convective zone is small, of the order of $P \sim  10^{-5}$. The reason for this low value is that the effective conductivity in the Sun and other stars is governed by radiative processes ~\cite{garaud2021journey}. In the Earth's liquid core, the Prandtl number is estimated to be of the order of $P \sim 0.1$ ~\cite{Thual1992zero}.  The magnetic Prandtl number in the Earth's outer core is $P_{m} \sim 10^{-6}$ ~\cite{Schaeffer2017turbulent}. For stellar interiors, the magnetic Prandtl number varies from $10^{-6}$ to $10^{-4}$, and for gas in the intergalactic medium, $P_{m} \gg  1$~\cite{Mondal2018onset}. The Rayleigh number in the convective zone is approximately $R  \sim 10^{20}$ and in the photosphere it is approximately $R \sim 10^{16}$~\cite{OSSENDRIJVER2003}. The Taylor number in the convective zone of the Sun is estimated to be of the order of $Ta \sim  10^{27}$ ~\cite{OSSENDRIJVER2003}. In our simulations, we employ values for these control parameters with different orders of magnitude, for numerical reasons and because we are interested in identifying the first routes to chaos and the onset of dynamo.

We assumed stress-free perfectly electrically conducting horizontal boundaries, i.e., at $z = 0$ and $z=1$ the following holds: 
\begin{equation}
\frac{\partial v_x}{\partial z} = \frac{\partial v_y }{ \partial z} =  v_z = 0,
\label{eqn:6}
\end{equation}
\begin{equation}
\frac{\partial b_x}{\partial z} = \frac{\partial b_y }{ \partial z} =  b_z = 0, 
\label{eqn:7}
\end{equation}
and $\theta=0.$
Periodicity in  horizontal directions with the same period $\emph{L}$ is assumed,
\begin{equation*}
\mathbf{v}(x, y, z)  = \mathbf{v}(x + mL, y + nL, z),
\end{equation*} 
\begin{equation}
\mathbf{B}(x, y, z)  = \mathbf{B}(x + mL, y + nL, z), 
\label{eqn:per}
\end{equation} 
\begin{equation*}
\theta(x, y, z)   = \theta(x + mL, y + nL, z),
\end{equation*} 
for all $m, n \in \mathbb{Z}$.

Equations  (\ref{eqn:1})--(\ref{eqn:3}) are solved numerically by applying the pseudospectral method~\cite{Canuto2007spectral}. The fields are represented as a truncated Fourier series satisfying the boundary conditions (\ref{eqn:6})--(\ref{eqn:per}):
\begin{equation}
{\bf v} = \sum_{n_1=-N/2+1}^{N/2-1}\sum_{n_2=-N/2+1}^{N/2-1}\sum_{n3=0}^{N/2-1}
\begin{pmatrix}
  \mathit{\hat{v}^{x}_\mathbf{n}} \textrm{cos}(\pi n_{3}z) \\ 
  \mathit{\hat{v}^{y}_\mathbf{n}} \textrm{cos}(\pi n_{3}z)\\ 
  \mathit{\hat{v}^{z}_\mathbf{n}} \textrm{sin}(\pi n_{3}z) 
\end{pmatrix} 
\exp\left(2\pi i(n_{1}x+n_{2}y)/L\right), 
\label{eqn:vf}
\end{equation}
${\bf B}$ is expanded analogously and $\theta$ is expanded as $v_z$. In computations we used multidimensional forward and backward fast Fourier transforms implemented in the library FFTW \cite{FFTW}. In simulations of both hydrodynamic and magnetohydrodynamic attractors, with the resolution of $N=48$ and $N=64$, respectively, our tests showed no significant influence of removal of aliasing errors, so most of the computations were performed without dealiasing.  

The system of ordinary differential equations for the Fourier coefficients, resulting upon substitution of the Fourier series to the governing equations (\ref{eqn:1})--(\ref{eqn:4}), is solved using the third-order exponential time-differencing method  ETDRK3 ~\cite{Cox2002exponential}.

\section{HYDRODYNAMIC SIMULATIONS} \label{section-3}

\subsection{NUMERICAL RESOLUTION} \label{Resolution hydrodynamic}

To choose a numerical resolution for the purely hydrodynamic simulations (${\bf B}=0$), we compute the spatial kinetic energy spectra for different numerical resolutions and $L=4$ and $P =0.3$. 

\begin{figure}[th!]\begin{center}
\includegraphics[width=1.\columnwidth]{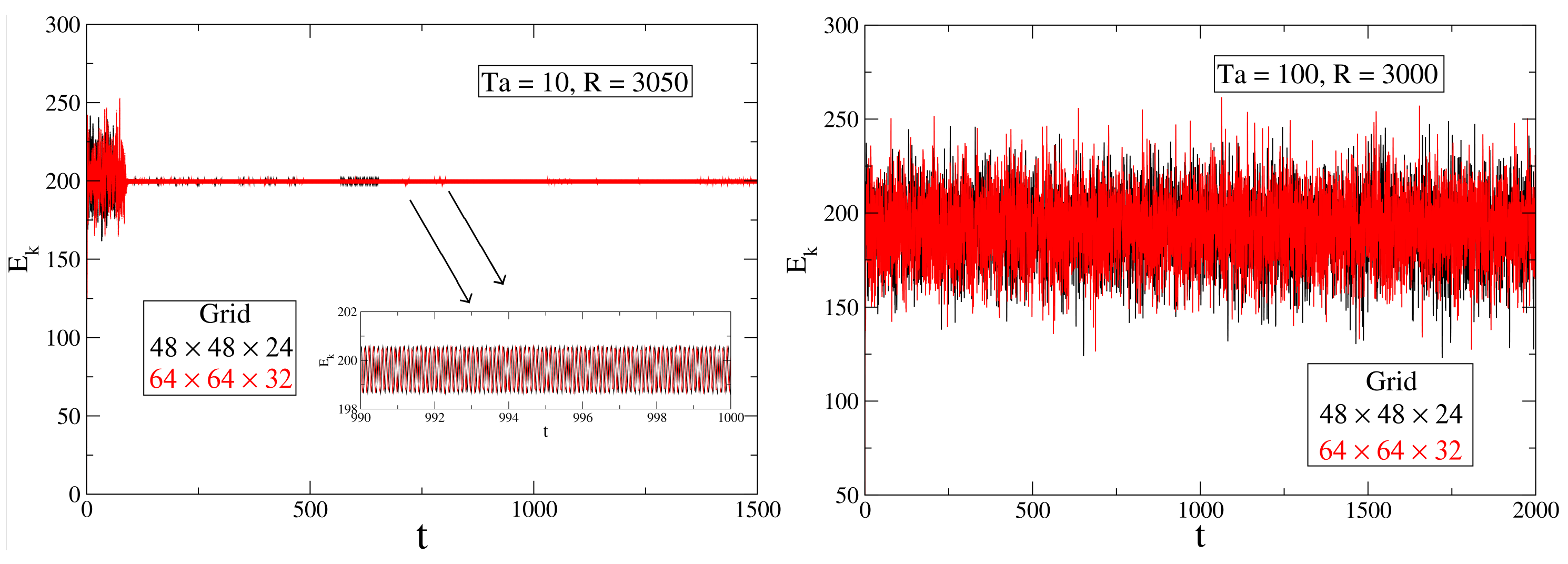}
\end{center}\caption[]{
Kinetic energy time series for the RBC model with $L=4$, $P=0.3$, $R = 3000$ and $Ta = 100$ (left panel) and $R = 3000$ and $Ta = 100$ (right panel). 
The black line represents computations with spatial resolution of $48 \times 48 \times 24$ and the red line with $64 \times 64 \times 32$.
}
\label{fig_1}\end{figure}

Figure \ref{fig_1} shows a comparison of the time series of kinetic energy for two different numerical resolutions and two dynamically different regimes.
In the left panel, the time series for $Ta=10$ and $R=3050$ are shown for $N=48$ in (\ref{eqn:vf}) (i.e., $48 \times 48 \times 24$ Fourier harmonics in total; black line) and $N=64$ (i.e., $64\times 64 \times 32$ harmonics; red line).  
Both curves overlap and both states are qualitatively the same, exhibiting an initial transient with strong chaos followed by periodic oscillations. In the right panel, the time series for $Ta=100$ and $R=3000$ are shown, with both resolutions converging to the same chaotic attractor. Increasing the numerical resolution leads to a significant increase in the computational time, therefore, since we need to compute many long-time runs, we choose the spatial resolution of $48 \times 48 \times 24$ Fourier harmonics for hydrodynamic simulations. We have verified that this resolution is sufficient to identify the attractors mentioned in Podvigina \cite{Podvigina2008magnetic}, where several attractors and bifurcations of the same convective system were identified.

\subsection{HYDRODYNAMIC CONVECTIVE ATTRACTORS}

Following Podvigina \cite{Podvigina2008magnetic}, where equations (\ref{eqn:1})-(\ref{eqn:4}) were studied without rotation, 
we investigate the hydrodynamic convective attractors by adding rotation. In this section, we provide a description of convective regimes in the absence of a magnetic field as a function of $R$ and $Ta$. In all simulations the Prandtl number was fixed at $P = 0.3$ and a square convective cell with $L=4$ was considered. For each set of control parameter values ($R$ and $Ta$), random initial conditions were used for the velocity fields
 and integration was conducted until an attractor was reached. The results are summarized in table \ref{tab:table_1}, which distinguishes the temporal behavior of the hydrodynamic attractors as the trivial steady states $\textbf{v}=0$ (-) and non-trivial ones $|\textbf{v}| > 0$  (S), periodic (P), quasiperiodic (QP), and chaotic (C) convection. For each value of $R$, 
as the Taylor number is increased, we observe a sequence of QP, P, and C regimes, until the state with no motion for large enough $Ta$ (i.e., where convection is suppressed by the Coriolis force) is reached. For fixed $Ta$, an increase in $R$ also leads to a complex sequence of P, QP, and C attractors, but for higher values of $Ta$ chaotic convective attractors are more frequent. 

As $R$ is increased, the critical value of $Ta$ (corresponding to the marginal stability of the trivial steady state) also increases. The onset of convection can be studied analytically as a linear stability problem of the motionless state (see Chandrasekhar \cite{chandrasekhar1961hydrodynamic}) where marginal stability boundary on the Rayleigh and Taylor numbers plane and critical wave numbers were determined. However, no closed form of the stability curve was presented (critical wave numbers are solutions to a cubic equation), but for some values of $Ta$ the critical Rayleigh and horizontal wave numbers were tabulated (see Table VII {\it ibid}.). Later, in Kloosterziel and Carnevale \cite{kloos}, the closed formulae were derived for rotating Rayleigh-B\'enard convection with free-free horizontal boundaries considered in this paper. In the absence of rotation, convection sets in at $Ra\approx657$.

\begin{table}[t]
\begin{center}
    \caption{\label{tab:table_1}Attractors of the convective hydrodynamic system as a function of $R$ and $Ta$  for $P = 0.3$. The dashes indicate the absence of motion ($\textbf{v}=0$); S stands for steady states; P, periodic; QP, quasiperiodic, and C are chaotic attractors.}
    \footnotesize
\begin{tabular}{@{}llllllllllllllll}
\br
R/Ta & 0 & 2 & 5 & 10 & 50 & 100 & 400 & 600 & 1000 & 1500 & 1600 & 2000 & 2500 & 3000 & 3600\\ 
\mr
700  & S  & S  & S  & S  & -  & -  & - & - & - & - & - & - & -  & -  & -\\
800  & C  & C  & QP & C  & -  & -  & - & - & - & - & - & - & -  & -  & -\\
900  & C  & P  & C  & C  & P  & P  & - & - & - & - & - & - & -  & -  & -\\
1000 & QP & P  & C  & C  & C  & C  & - & - & - & - & - & - & -  & -  & -\\
1100 & QP & QP & P  & P  & C  & C  & - & - & - & - & - & - & -  & -  & -\\
1200 & QP & QP & P  & P  & P  & C  & C & - & - & - & - & - & -  & -  & -\\
1300 & QP & QP & P  & P  & P  & C  & C & - & - & - & - & - & -  & -  & -\\
1400 & QP & QP & P  & P  & P  & C  & C & C & - & - & - & - & -  & -  & -\\
1500 & QP & C  & P  & P  & P  & C  & C & C & - & - & - & - & -  & -  & -\\
1600 & QP & C  & P  & P  & P  & QP & C & C & - & - & - & - & -  & -  & -\\
1700 & QP & C  & QP & C  & P  & QP & C & C & C & - & - & - & -  & -  & -\\
1800 & QP & QP & C  & QP & P  & QP & C & C & C & - & - & - & -  & -  & -\\
1900 & QP & QP & QP & C  & P  & QP & C & C & C & - & - & - & -  & -  & -\\
2000 & QP & C  & QP & C  & QP & P  & C & C & C & - & - & - & -  & -  & -\\
2100 & QP & C  & C  & C  & C  & QP & C & C & C & C & - & - & -  & -  & -\\
2200 & C  & C  & C  & QP & C  & QP & C & C & C & C & C & - & -  & -  & -\\
2300 & C  & C  & C  & P  & C  & QP & C & C & C & C & C & C & -  & -  & -\\
2400 & C  & C  & P  & P  & C  & QP & C & C & C & C & C & C & QP & -  & -\\
2500 & C  & QP & P  & P  & C  & C  & C & C & C & C & C & C & C  & QP & -\\
2600 & C  & QP & P  & P  & C  & C  & C & C & C & C & C & C & C  & C  & C\\
2700 & C  & QP & QP & P  & C  & C  & C & C & C & C & C & C & C  & C  & C\\
2800 & C  & QP & QP & P  & C  & C  & C & C & C & C & C & C & C  & C  & C\\
2900 & C  & QP & QP & P  & C  & C  & C & C & C & C & C & C & C  & C  & C\\
3000 & C  & C  & QP & P  & C  & C  & C & C & C & C & C & C & C  & C  & C\\
3100 & C  & C  & C  & QP & C  & C  & C & C & C & C & C & C & C  & C  & C\\
3200 & C  & C  & C  & QP & C  & C  & C & C & C & C & C & C & C  & C  & C\\
3300 & C  & C  & C  & QP & C  & C  & C & C & C & C & C & C & C  & C  & C\\
3400 & C  & C  & C  & C  & C  & C  & C & C & C & C & C & C & C  & C  & C\\
3500 & C  & C  & C  & C  & C  & C  & C & C & C & C & C & C & C  & C  & C\\
\br
\end{tabular}\\
\end{center}
\end{table}
\normalsize

Attractors described in table 1 provide a preliminary information on attractors of the convective system when two parameters are varied. In order to study the transitions to chaos in more details, we construct a set of bifurcation diagrams for three particular values of $Ta$ (10, 50 and 100) and varying $R$ as the control parameter, $R\in[700,4000]$. Bifurcation diagram for $Ta=0$ is presented in \cite{Podvigina2008magnetic} and transition to hyperchaos is detailed in \cite{Chertovskih2015route}. For these values of $Ta$, we investigate attractors of the system and bifurcations delimiting branches of the attractors. A Poincaré map was adopted by selecting only the local maximum points of the kinetic energy. 

The non-rotating case ($Ta=0$, attractors presented in the second column of table~\ref{tab:table_1}) is studied in detail in \cite{Podvigina2008magnetic} (see Section 6.1 {\it ibid.}) in a shorter range of the control parameter, for $R\le2500$. A complex sequence of steady, periodic, quasiperiodic and chaotic attractors is found. In total, 14 branches of attractors are found: 3 steady states, 5 periodic, 4 quasiperiodic, and 2 chaotic regimes.
The coexistence of attractors is also observed. 
Due to coarse resolution in $R$, the convective attractors for $Ta=0$ shown in table 1, do not represent all branches found in \cite{Podvigina2008magnetic}, however, we checked that our results are in agreement.  
Below we discuss what are the main differences of the convective attractors in the non-rotating system and in the presence of rotation.

\begin{figure}[th!]\begin{center}
\includegraphics[width=0.7\columnwidth]{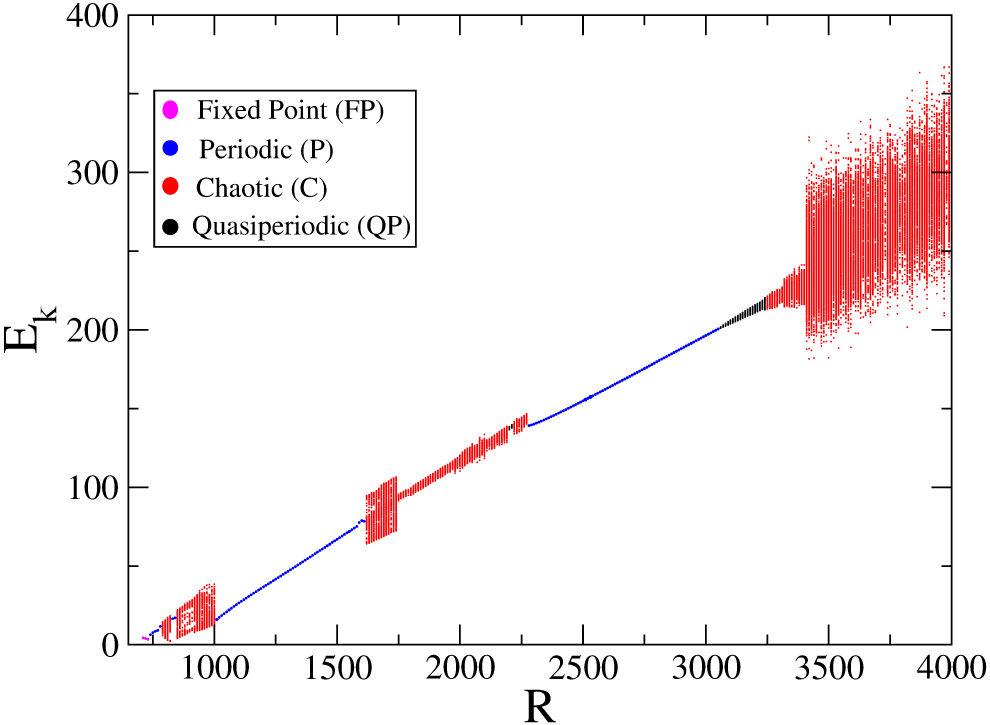}
\end{center}\caption[]{
Bifurcation diagram of the kinetic energy as a function of  $R$, for $Ta = 10$. 
}\label{fig_2}\end{figure}

For the bifurcation diagram in figure~\ref{fig_2} we fixed the Taylor number at $Ta = 10$ and varied the Rayleigh number up to $R = 4000$. The bifurcation diagram was computed using the continuation in parameter technique: {\it i}) A random initial condition is chosen for the initial value of $R$, then the system equations are solved and the initial transient is dropped; {\it ii}) the kinetic energy of the 
 attractor is plotted for a few hundred Poincaré points; {\it iii}) $R$ is increased and the final system state from the previous $R$ value is used as the initial condition for the run with the new $R$ value and the integration and plotting procedure is repeated. Attractors of the same temporal nature obtained by this procedure are assumed to constitute the same branch of attractors. 
The convective system initially (for small supercritical values of $R$ for convection) exhibits steady-state, or fixed-point behavior, represented by magenta circles. At $R=R_c$ ($730<R_c<740$), a first Hopf bifurcation takes place, where a time-periodic state is emerging,  shown by blue circles. A period-doubling bifurcation occurs in $780<R<780.96$. Next, the system becomes chaotic and remains so until $R = 1010$, when the chaotic regime looses its stability and the chaotic attractor disappears. This seems to be an embedded saddle-node bifurcation, but we did not pursue the saddle point. 
The behavior remains periodic until it 
switches to a chaotic state. This bifurcation occurs at $1610 < R < 1620$. The chaotic behavior persists until $R = 2190$.

Within the interval $2190\le R\le2200$, the system displays quasiperiodic behavior,  represented by black circles in the bifurcation diagram. In the range $2210 \le R \le 2220$, the attractor becomes chaotic again. Within $2270 \le R \le 2280$ the system switches back to a periodic attractor. This periodic behavior continues until the interval of $3050 \le R \le 3060$, after which another Hopf bifurcation occurs, leading 
 the system back to quasiperiodic behavior. Within the interval $3240 < R < 3250$, the branch of quasiperiodic attractors loses its stability and a new chaotic attractor is formed. This chaotic behavior persists until the largest value we considered, $R=4000$.

\begin{figure*}[th!]\begin{center}
\includegraphics[width=1\columnwidth]{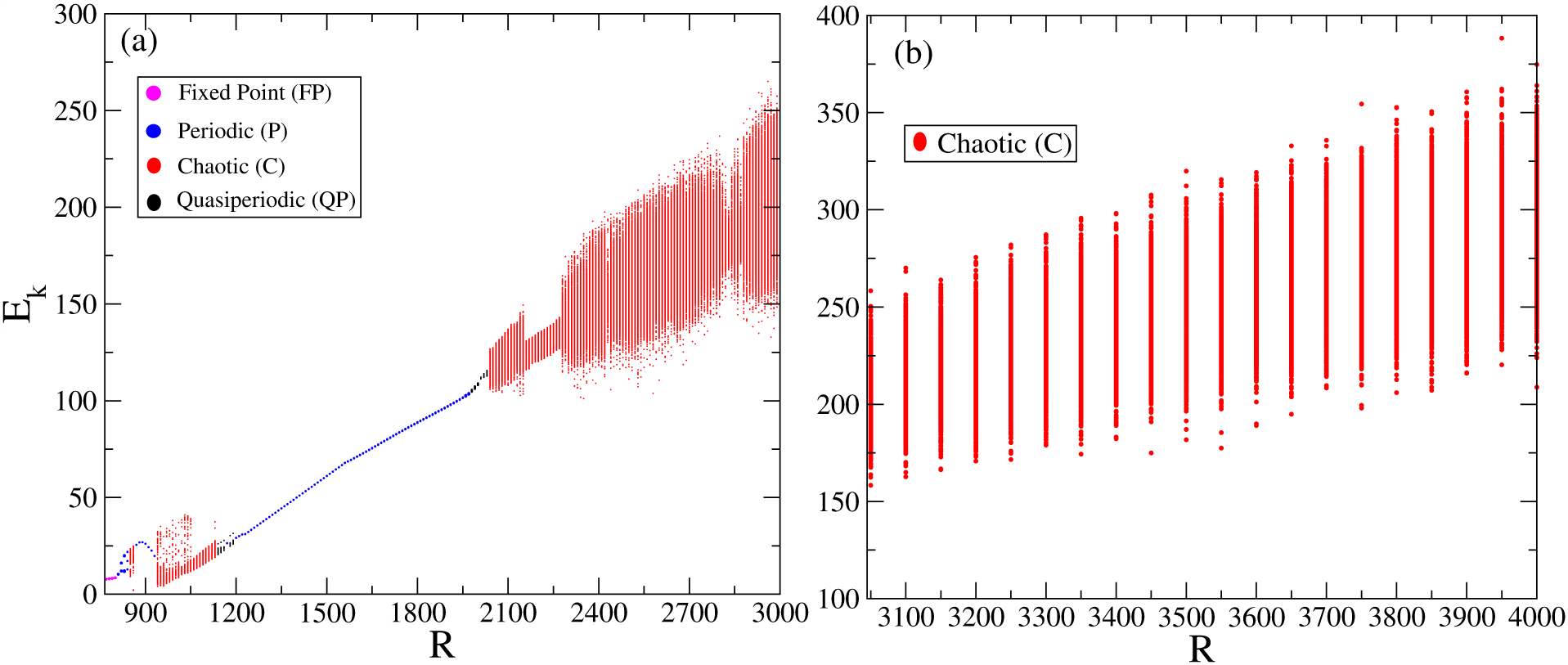}
\end{center}\caption[]{
Bifurcation diagram of the kinetic energy as a function of  $R$, for $Ta = 50$. 
}
\label{fig_3}\end{figure*}

For $Ta = 50$, the results are detailed in the bifurcation diagram shown in figure~\ref{fig_3}.
The initial sequence of bifurcations is somehow similar to what is presented in figure~\ref{fig_2}, although some bifurcation points are shifted to the right in $R$. The fixed-point attractor bifurcates to a periodic orbit, then to quasiperiodic and chaos as prescribed by the typical Ruelle-Takens-Newhouse route to chaos \cite{ruelle,newhouse}. 
In the $2040<R<2050$ range, a QP attractor loses stability, giving rise to a chaotic attractor that persists until the end of the bifurcation diagram at $R=4000$. 

\begin{figure*}[th!]\begin{center}
\includegraphics[width=\columnwidth]{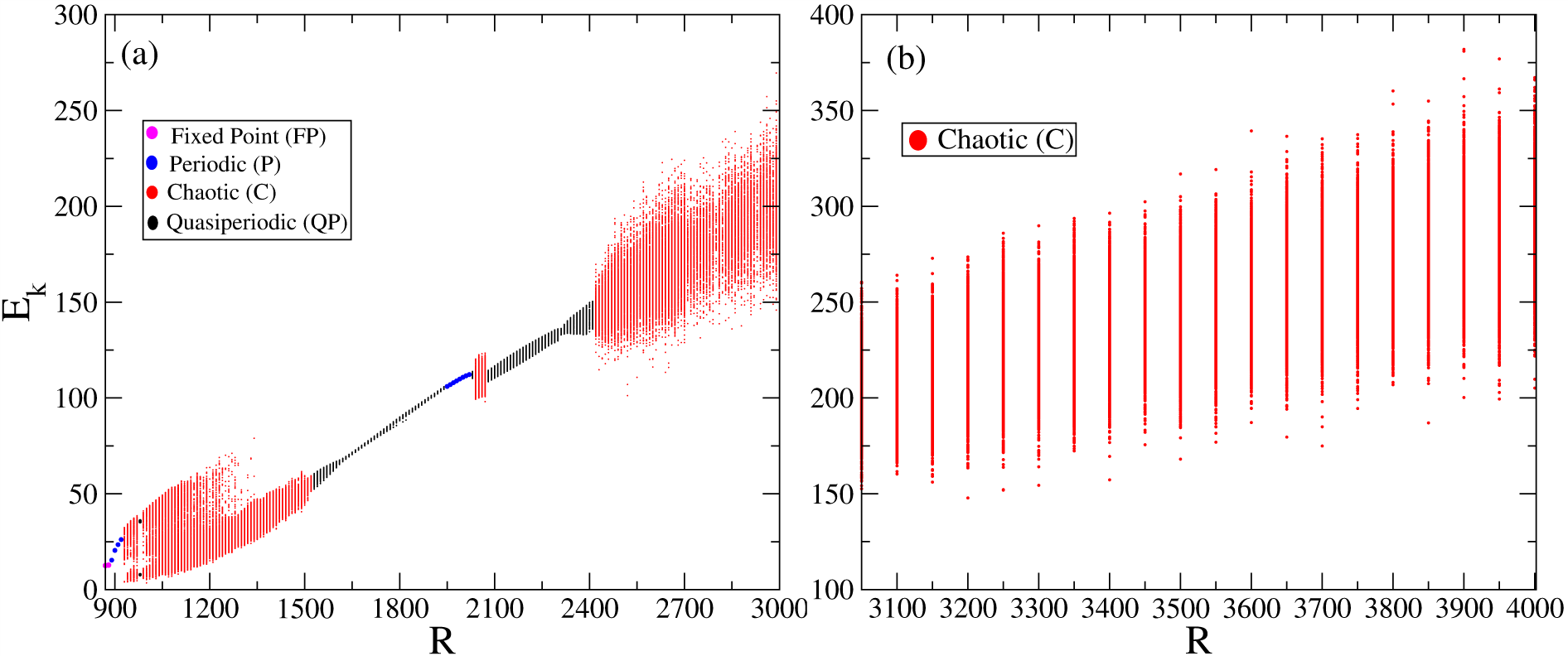}
\end{center}\caption[]{
Bifurcation diagram of the kinetic energy as a function of  $R$, for $Ta = 100$. 
} \label{fig_4}\end{figure*}

The bifurcation diagram in figure~\ref{fig_4} is for $Ta=100$. It also begins with a Ruelle-Takens-Newhouse route to chaos, but branches of time-periodic states are shorter (in $R$) for larger values of $Ta$, and QP and C states are stable in larger intervals. A common feature of all three diagrams (figures~\ref{fig_2}, \ref{fig_3}, and \ref{fig_4}) is a complex sequence of Hopf, period-doubling and saddle-node bifurcations for low $R$, the absence of chaos for intermediate values of $R$, and a transition to chaos via quasiperiodicity for higher values of $R$.

Chertovskih et al. \cite{Chertovskih2010} numerically investigated the hydrodynamic regimes ($\mathbf{B = 0}$), varying the Taylor number from $0$ to $2000$, and kept the other parameters fixed. They observed a typical bifurcation sequence, finding 5 types of hydrodynamic attractors. The results indicate that the critical Rayleigh number for the initiation of convection grows with the Taylor number, and the rapid rotation rate in hydrodynamic convection disrupts convective flows. These results are in agreement to ours. 

\begin{figure*}[th!]\begin{center}
\includegraphics[width=\columnwidth]{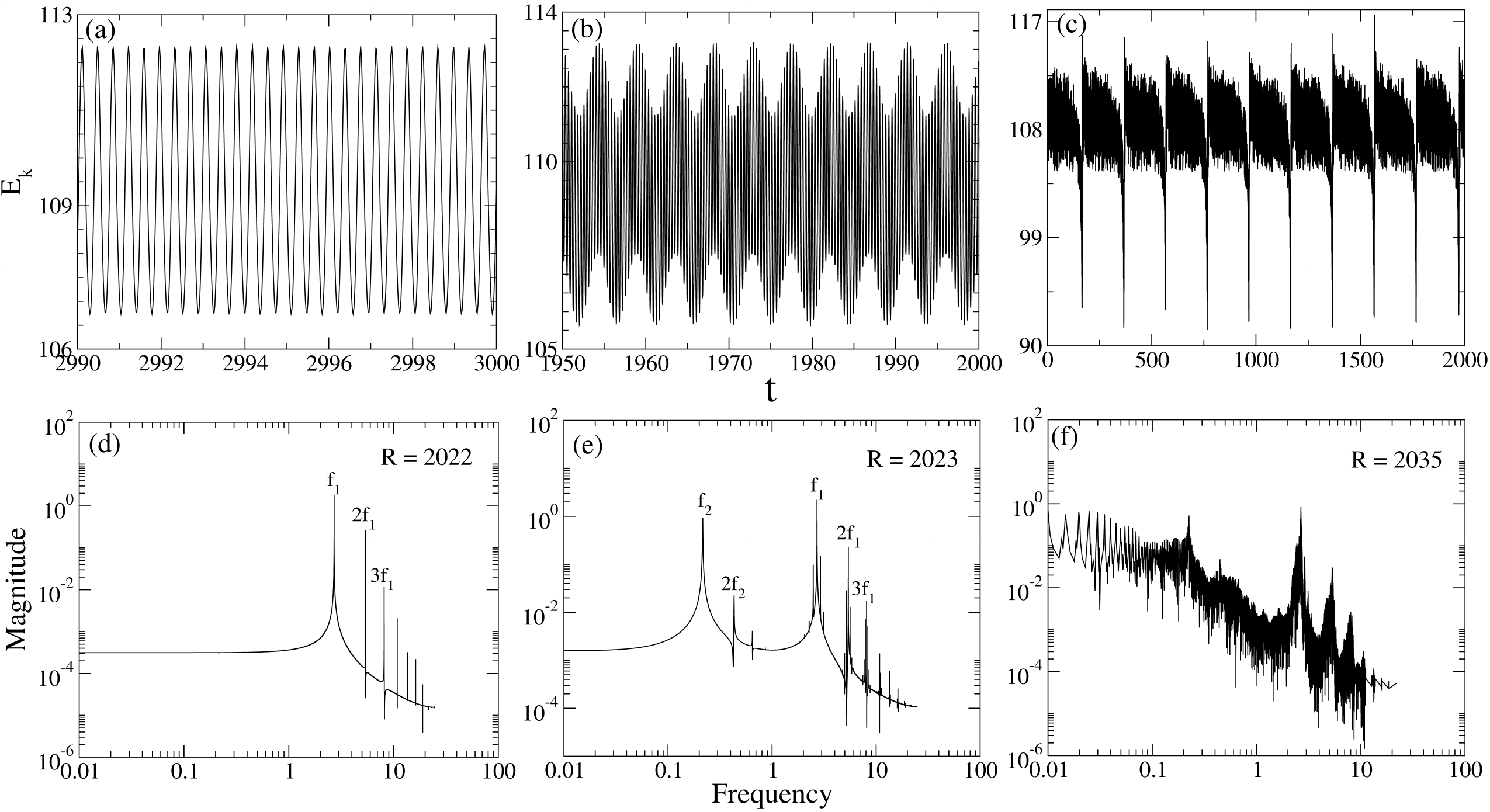}
\end{center}\caption[]{
Time series of the kinetic energy (top panel) for $Ta = 100$ and (a) $R = 2022$, (b) $R = 2023$, (c) $R = 2035$. Frequency spectra (bottom panels) are computed for the corresponding time series of the kinetic energy. In (a) the attractor is periodic, (b) quasiperiodic with two incommensurate frequencies, and in (c) it is chaotic.}\label{fig_5}\end{figure*}

The classification of the attractors in the previous diagrams was done by analyzing the power spectra of the kinetic energy time series \cite{Stoica2005spectral}. Examples of kinetic energy time series and frequency spectra are shown in figure \ref{fig_5}.
The upper panel shows the time series of kinetic energy, while the frequency spectra are shown in the lower panel for different values of $R$. In particular, figure \ref{fig_5}(d) shows the frequency spectrum for a periodic attractor at $R = 2022$, possessing a fundamental frequency $f_{1} \approx 2.70828$ and its higher harmonics, $2f_{1} \approx 5.41656$ and $3f_{1} \approx 8.12484$. If $R$ is slightly increased, $R=2023$, we observe the emergence of a new frequency, $f_2$, incommensurate with $f_1$. In panel (e) the two basic time frequencies are $f_{1} \approx 2.70271$ and $f_{2} \approx 0.216216$, hence, the attractor is quasiperiodic. If $R$ is increased further, the system undergoes a bifurcation (or a sequence of bifurcations) giving rise to a chaotic attractor. The temporal spectra of the chaotic attractor is dense (see, e.g., for $R=2035$ in figure~\ref{fig_5}(f)). The sequence: periodic, quasiperiodic and then chaotic state, is the standard Ruelle-Takens-Newhouse route to chaos.

Comparing the attractors in the rotating system described above and the convective regimes in the absence of rotation reported in \cite{Podvigina2008magnetic}, we conclude that the influence of rotation on convection is non-monotonic and in no way simple. For all values of $Ta$, the system becomes chaotic when $R$ is large enough. However, the interval of the existence of the branch of chaotic states is very sensitive to the rotation rate: In the non-rotating system the chaotic regimes are observed at  $R\ge1990$; in the rotating system, for $Ta=10$ the chaotic states are found at $R\ge3250$; for $Ta=50$, at $R\ge2050$ and, finally, for $Ta=100$ at $R>2413$. 
Also, there are similarities in the qualitative behaviour of the convective system. For instance, there is a branch of chaotic attractors near the onset of convection (for small values of $R$), however, the intervals of their stability differ significantly: 
for $Ta=0$, chaotic attractors are found in $740\le R\le995$;  
for $Ta=10$, in $790\le R\le1010$; 
for $Ta=50$, in $850\le R\le870$, and 
for $Ta=100$, in $930\le R\le985$. 
For a fixed value of $R$, analysis of table~\ref{tab:table_1} let us conclude that most regimes of convection in the rotating systems become more ``chaotic'' when the rotation rate is higher
not only for vigorous convection, but also close to the onset. However, for moderate values of $R$, in the absence of rotation, the longest branches are quasiperiodic, for $Ta=10$ and 50 they are periodic (see blue dots in figures~\ref{fig_2} and \ref{fig_3}, and for $Ta=100$ are quasiperiodic again (see black dots in figure~\ref{fig_4}).

We also checked if an intermittent switching between chaotic and quasiperiodic phases takes place in the rotating system as it does in the absence of rotation~\cite{Chertovskih2015route}. Such intermittency is found to be important for magnetic field generation~\cite{Chertovskih2017}, because different phases of the intermittency have different ability to be a dynamo (the quasiperiodic phases are found to be more beneficial for dynamo than the chaotic ones). We found that even for a very weak rotation (the smallest value considered was $Ta=0.1$) the rotating systems does not display intermittency in the parameter range found in~\cite{Chertovskih2015route}. This difference can be related to the fact that the presence of the Coriolis term significantly affects the symmetry group of the dynamical system (it breaks the reflections, see~\cite{Chertovskih2010} for details), changing drastically the geometry of the phase space. 

\section{MAGNETOHYDRODYNAMIC SIMULATIONS} \label{section-4}
\subsection{NUMERICAL RESOLUTION}

\begin{figure*}[th!]
  \centerline{\includegraphics[width=\columnwidth]{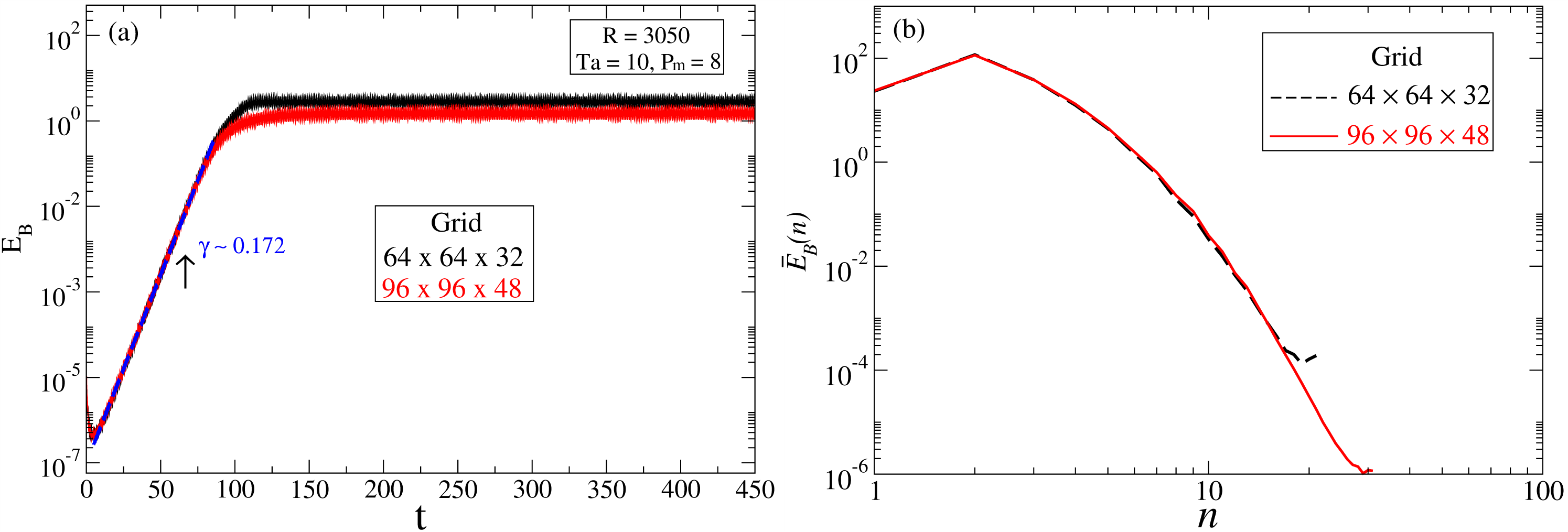}}
  \caption{(a) Comparison of log-linear scale magnetic energy time series with different numerical resolutions, for $R = 3050$, $Ta = 10$, $P=0.3$, and $P_{m} = 8$. The black line corresponds to a numerical resolution of $64 \times 64 \times 32$ Fourier harmonics; the red line corresponds to $96 \times 96 \times 48$ harmonics. In the kinematic regime the growth rate is $\gamma \approx 0.172$ for both resolutions (blue slope). (b) Time-averaged magnetic energy spectra corresponding to the spatial resolutions of $64 \times 64 \times 32$ (dashed black line) and $96 \times 96 \times 48$ points (solid red line).}
\label{resol-numer}
\end{figure*}

Simulations of evolution of magnetic field in (\ref{eqn:1})-(\ref{eqn:4}) usually demand a higher numerical resolution than the purely hydrodynamic case, because dynamos considered in this work operate for $P_m>1$, hence, diffusion of the magnetic field takes place on smaller spatial scales than for fluid flow (cf. coefficients with the Laplacians in the Navier-Stokes equation (\ref{eqn:1}) and in the magnetic induction equation (\ref{eqn:2})). Therefore, we repeat the resolution test (described in section~\ref{Resolution hydrodynamic}) adding a weak magnetic field to the initial conditions. The initial magnetic field is $\mathbf{B}(\mathbf{x},0) = (\cos(\pi x_2 / 2), 0,0)$, scaled so that the magnetic energy is $E_b(0) = 10^{-5}$.
Figure \ref{resol-numer}(a) shows a comparison of the time series of the magnetic energy on log-linear scales for RBC MHD simulations using $64 \times 64 \times 32$ (black line) and $96 \times 96 \times 48$ (red line) Fourier harmonics, i.e., resolutions for $N=64$ and $N=96$ in the truncated Fourier series (\ref{eqn:vf}) are considered. In both cases, we use a periodic hydrodynamic convective attractor as the initial condition for the velocity field and a supercritical magnetic Prandtl number (i.e., enabling the dynamo to operate). The following values of the control parameters are adopted: $P_{m} = 8$, $R = 3050$, $Ta = 10$, and $P = 0.3$. The growth rate $\gamma$ in the kinematic phase is approximately $ \gamma \sim 0.172$ for both numerical resolutions.
In the saturated regime, the attractors of the MHD system are chaotic in both cases and exhibit similar time-averaged magnetic energy spectra, as seen in figure \ref{resol-numer}(b), where the dashed black line corresponds to $64\times 64\times 32$ Fourier harmonics and the solid red line corresponds to $96\times 96\times 48$ harmonics. Both attractors are identical qualitatively, however, the magnetic energy of the attractor for the higher resolution is slightly smaller, because the magnetic diffusion is better resolved. Based on this, we chose to perform simulations at a lower numerical resolution of $64 \times 64 \times 32$, which is less expensive computationally (since we plan to perform long simulations to achieve attractors of the dynamical system), and because we are interested in qualitative analysis of the processes in the MHD system under investigation.

\subsection{ONSET OF DYNAMO ACTION}

In this section, we fix $P = 0.3$ and $R=3000$, and consider several convective attractors (described above) for different values of the Taylor number from 0 (no rotation) to 2500. For these convective attractors we perform a numerical study aiming to estimate $P_m^c$, the critical value of $P_m$ where the onset of dynamo action occurs (i.e., for $P_m<P_m^c$ a small initial magnetic field decays, while for $P_m>P_m^c$ it grows exponentially). We consider values of $P_m$ ranging from 1 to 10, with a step size of 1. The periodic, quasiperiodic, and chaotic hydrodynamic convective attractors obtained in section \ref{section-3} were used to study their ability to generate magnetic field in the kinematic and nonlinear regimes. 
The seed magnetic field is $\mathbf{B}(\mathbf{x},0) = (\cos(\pi x_2 / 2), 0,0)$, scaled so that the magnetic energy is $E_b(0) = 10^{-7}$.

\begin{figure*}[th!]\begin{center}
\includegraphics[width=0.9\columnwidth]{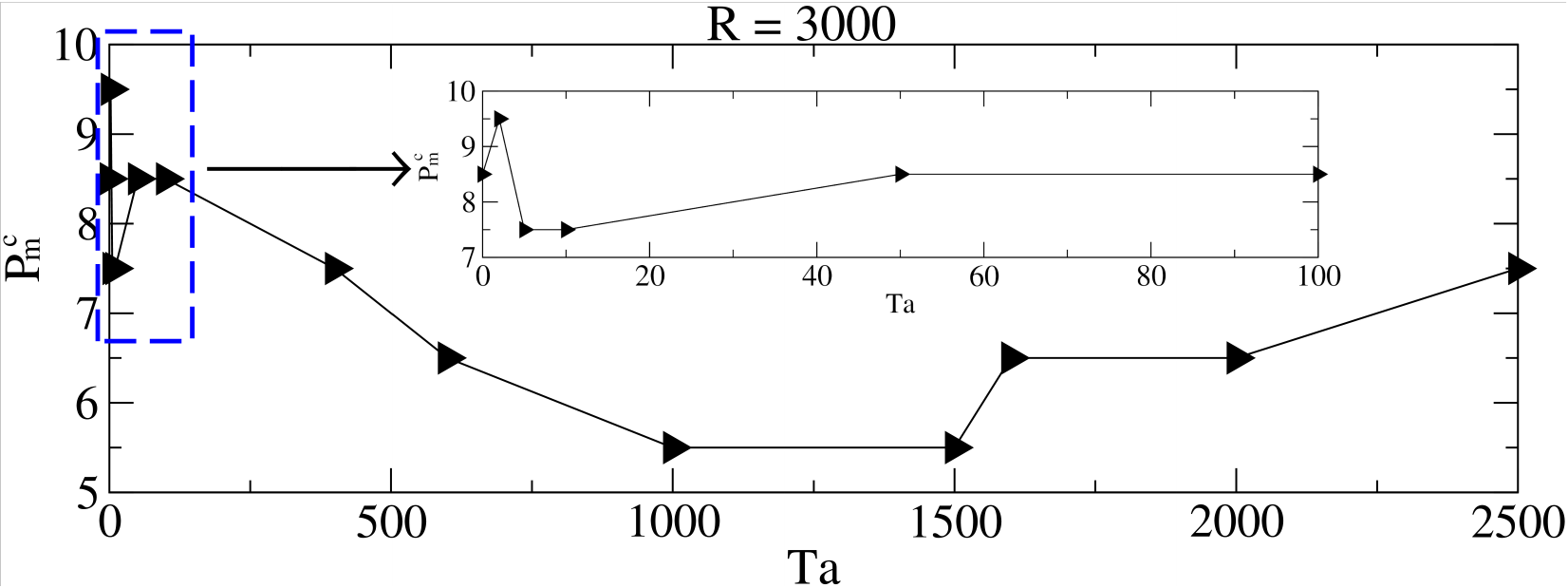}
\end{center}\caption[]{
Estimate for the critical magnetic Prandtl number, $P_{m}^{c}$, as a function of $Ta$ for $R = 3000$. The inset represents an amplification of the region in the blue dashed rectangle.}\label{Prandtl-magnetico-critico}\end{figure*}

Figure \ref{Prandtl-magnetico-critico} shows a plot of the critical magnetic Prandtl number as a function of $Ta$. 
First, we calculate $P^{c}_{m}$ considering a chaotic hydrodynamic regime for $Ta = 0$ and estimate it to be approximately 8.5 (more precisely, $P_m^c$ is in the range $8 < P^{c}_{m} < 9$). In the rotating system,
we see an initial oscillation in $P_m^c$ between 7.5 and 9.5 until $Ta=100$, when a gradual decrease takes place from $P_m^c \approx 8.5$ at $Ta=100$ to $P_m^c \approx 5.5$ at $Ta = 1500$. For larger values of $Ta$, $P^{c}_{m}$ increases, reaching a value of $7.5$ at $Ta = 2500$. 

\begin{figure*}[th!]\begin{center}
\includegraphics[width=\columnwidth]{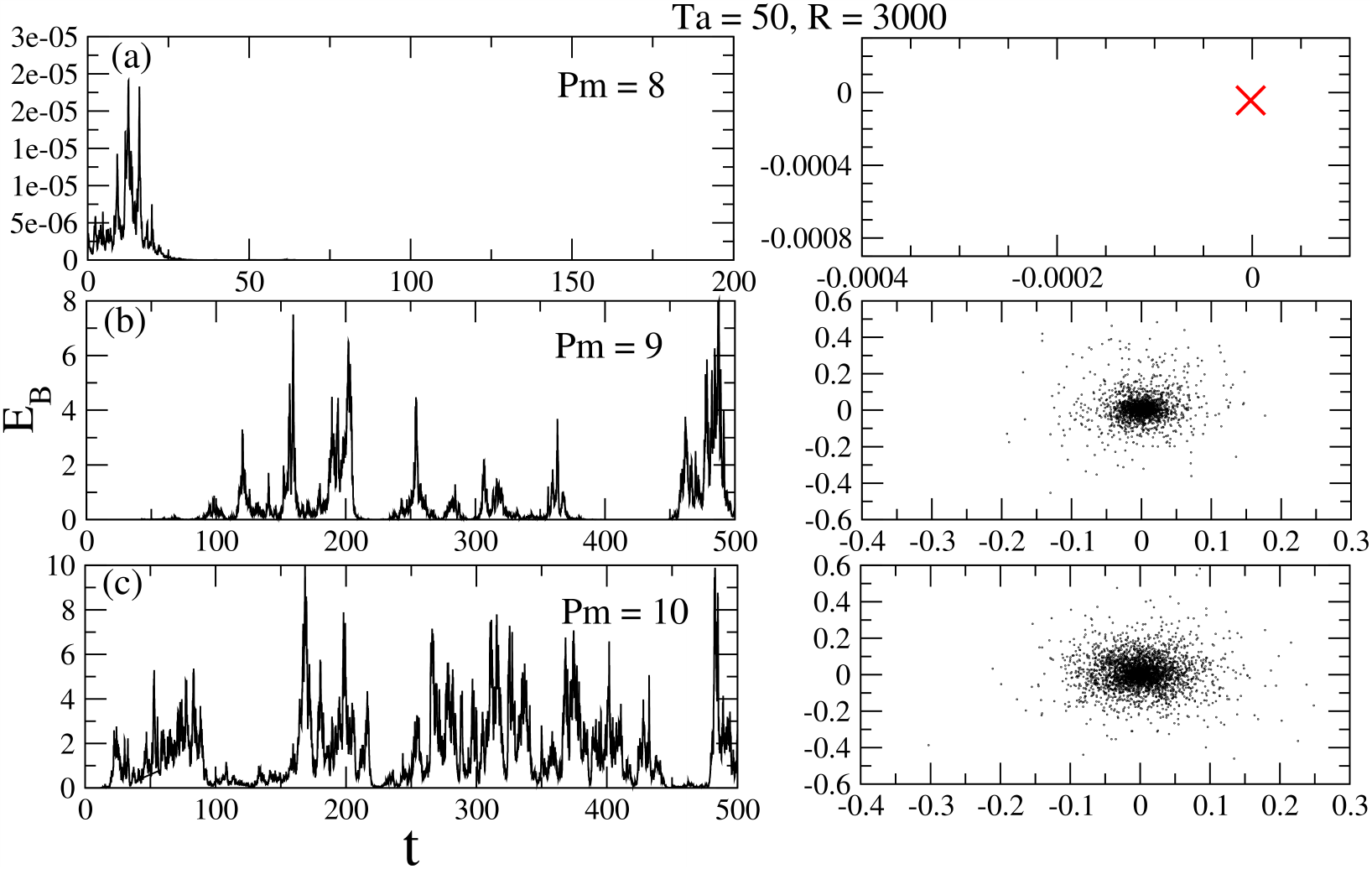}
\end{center}\caption[]{
Time series of the magnetic energy (left panel) and
projections (right plane) of the attractors onto the plane 
$Re(\hat{v}^{y}_{2,2,2})$ (horizontal axis), $Re(\hat{v}^{z}_{2,2,2})$ (vertical axis) for $Ta = 50$ and $R = 3000$. Before the transition to dynamo, the decay to a hydrodynamic attractor is observed in (a), while immediately after the transition, the magnetohydrodynamic attractors in (b) and (c) exhibit on-off intermittency.} \label{Ta=50-Pm=8-9-10}\end{figure*}

Figure \ref{Ta=50-Pm=8-9-10} illustrates the dynamics near the onset of dynamo. The left panels show the time series of the magnetic energy in the runs corresponding to three values of $P_m$ (8, 9 and 10), the right panels display the corresponding magnetohydrodynamic attractors projected onto a Poincare plane. We choose the plane defined by the real part of two components of a certain Fourier coefficient of the flow). We checked that other choices of the Poincare plane present a similar (qualitatively) behavior. It is shown in figure \ref{Ta=50-Pm=8-9-10}(a), that before the transition (no dynamo), the initially oscillating magnetic field decays and the system has a purely hydrodynamic attractor $(\mathbf{B} = 0)$, which is represented by the red cross in the right panel of figure \ref{Ta=50-Pm=8-9-10}(a).  
 As in Rempel et al. \cite{rempel2009novel}, Karak et al. \cite{karak2015}, and Oliveira et al. \cite{oliveira2021chaotic}, this illustrates a transient dynamo, governed by a nonattracting chaotic set, 
 also known as chaotic saddle ~\cite{nusse88,franco2020chaotic}. In figures \ref{Ta=50-Pm=8-9-10}(b), for $P_m = 9$, and \ref{Ta=50-Pm=8-9-10}(c), for $P_m = 10$, the magnetic energy time series exhibits on-off intermittency, alternating between bursty and quiescent phases. The right panels of (b) and (c) show the corresponding magnetohydrodynamic chaotic attractors. 
 The term ``on-off intermittency'' was first employed by Platt et al. \cite{Platt1993off} to denote an aperiodic switching between static, or laminar, behavior and chaotic bursts of oscillation, with the solar sunspot cycle being pointed as an example of the phenomenon, where cyclic variations in solar activity turned off and remain near zero for long periods.

 The intermittent time series in figure \ref{Ta=50-Pm=8-9-10} suggest that transition to dynamo in RBC with rotation is caused by a nonhysteretic blowout bifurcation, just like in the simpler setup of the non-rotating ABC-flow with periodic boundary conditions of Sweet et al. \cite{sweet2001alowout,Sweet2001blowout} and Rempel et al. \cite{rempel2009novel}. In this type of bifurcation, the dynamical system has a smooth invariant manifold with a chaotic attractor. After a blowout bifurcation,  the manifold loses transversal stability and the chaotic set ceases to be an attractor. Shortly after the transition, solutions exhibit on-off intermittency, spending a significant amount of time very close to the manifold, interrupted by short periods of strong chaotic bursts, away from the manifold. After each burst, the trajectory returns to the neighborhood of the manifold, and the process repeats intermittently \cite{Ott1994blowout}. 
 
The average duration of the laminar phases between bursts, $\tau$, in on-off intermittency decreases with distance from the critical parameter value, $P^{c}_{m}$. According to Grebogi et al. \cite{Grebogi1987}, in crisis transitions to chaos $\tau$ follows the scaling law
\begin{equation}
\tau \sim (P_{m} - P^{c}_{m})^{\gamma}.
\end{equation}

\begin{figure} [th!]
  \centerline{\includegraphics[width=0.6\columnwidth]{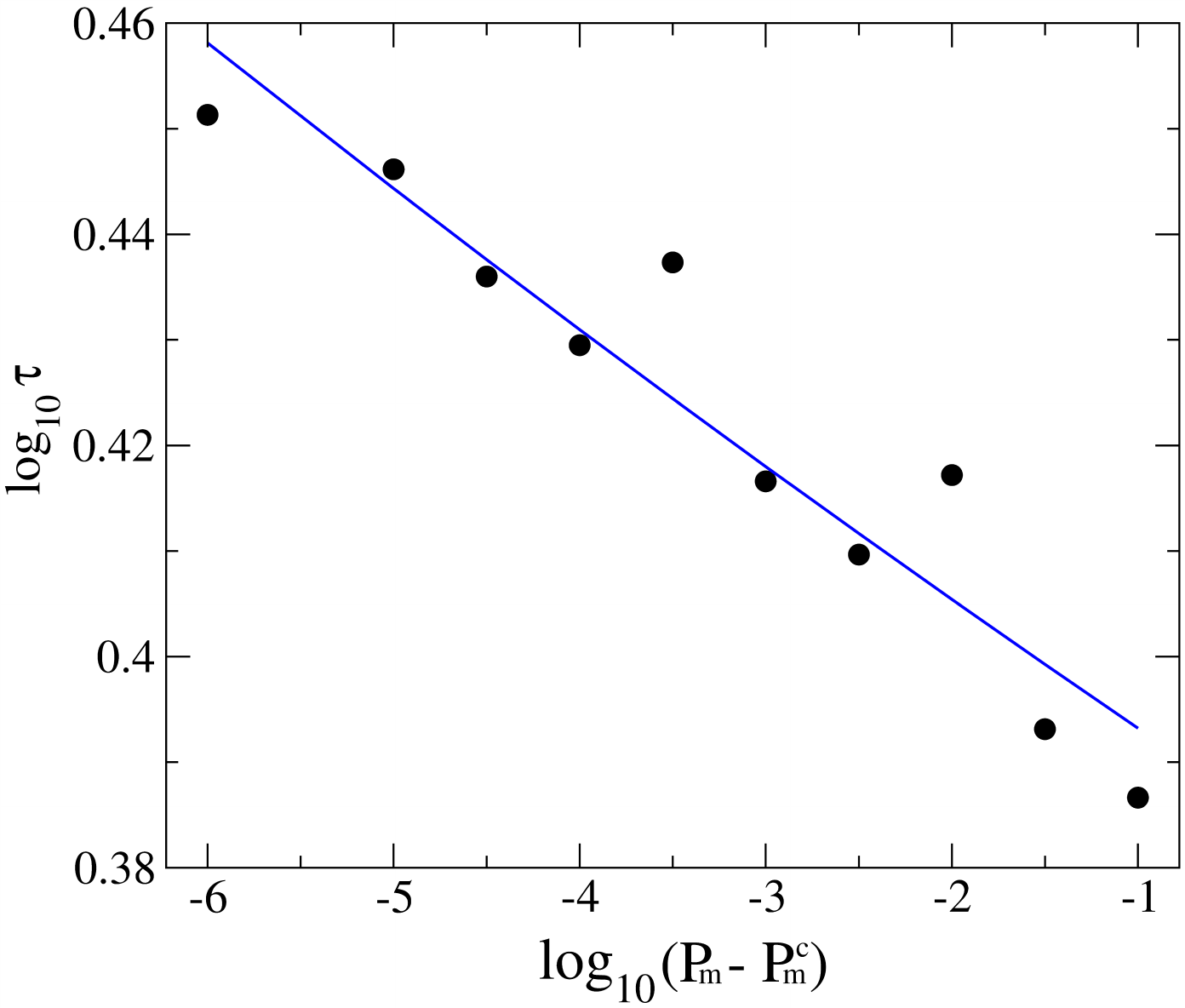}}
  \caption{Scaling law of the average duration of laminar phases between bursts $(\tau)$ as a function of distance to the critical value $(P_{m} - P^{c}_{m})$ at which intermittency appears, for $R = 3000$, $Ta = 100$ and $P_{m} = 9 > P^{c}_{m}$.}
\label{fig_11}
\end{figure}

We computed $\tau$ for a set of values of $P_{m}$ close to $P^{c}_{m}$ and obtained the results shown in figure \ref{fig_11}, where the fitted line has a slope of $\gamma  = -0.03$. The following procedure was adopted to construct this figure. First, a set of 100 initial conditions are selected from the chaotic intermittent hydromagnetic attractor at $P_{m} = 9 > P^{c}_{m}$, for $R = 3000$ and $Ta = 100$; 
then these initial conditions are used to generate long intermittent time series for different values of $P_m > P_m^c$; the value of $\tau$ is computed for each time series and an average $\tau$ is computed. Thus, each point in figure \ref{fig_11} is an average $\tau$ from 100 long time series. We consider that a chaotic burst starts whenever $E_{B}>2$. 
The fitted line was obtained by linear regression. This result endorses the conjecture that a nonhysteretic blowout bifurcation is responsible for the dynamo transition in rotating RBC.

\section{CONCLUSIONS} \label{section-5}

In this paper, we presented results from a numerical study of the transition to chaos in rotating RBC. Direct numerical simulations of the three-dimensional hydrodynamic and magnetohydrodynamic systems have been performed. In the hydrodynamic case, different routes to chaos were reported via quasiperiodicity. 
The critical Rayleigh number for the onset of convection grows with the Taylor number,  significantly impacting the onset of convection in the system. 
We also observed that rapid rotation interrupts convective flows, but if the Rayleigh number is large enough to produce convection, then increasing the Taylor number leads to more chaotic regimes than in the absence of rotation. 

From these studies of hydrodynamic convective regimes, we investigated the generation of a magnetic field by thermal convection. 
The on-off intermittency reported in our magnetic energy time series near dynamo onset shows similarity with the on-off intermittency reported by Sweet et al. \cite{Sweet2001blowout} and Rempel et al. \cite{rempel2009novel}. Both works considered the ABC-flow dynamo without rotation in a periodic box, where magnetic energy time series alternate between quiescent phases with magnetic energy near zero and strong energy burst phases. This type of intermittency occurs near critical values of the control parameters where a blowout bifurcation occurs. We obtained a scaling law for the quiescent phases as a function of the magnetic Prandtl number that confirms that a critical transition to dynamo of the nonhysteretic blowout type is found also in RBC. 

Childress and Soward (1972) ~\cite{childress1972convection} were the first to demonstrate that a fluid undergoing thermal convection in a rapidly rotating plane layer was capable of supporting a large-scale dynamo. Meneguzzi and Poquet ~\cite{meneguzzi1989turbulent} reported that in highly conductive fluids, both nonrotating and rotating thermal convection can aid the dynamo action, producing intermittent magnetic fields. Our results are similar to those of Meneguzzi and Pouquet ~\cite{meneguzzi1989turbulent}, where we found a transition to dynamo with rotation producing intermittent magnetic fields. Our investigation extends the results of Chertovskih et al.  ~\cite{Chertovskih2010}, who studied magnetic field generation in the rotating RBC system for $L=2\sqrt{2}$, $P=1$, $P_m=8$, $R=2300$ varying the Taylor number. By adopting $P=0.3$ and various $R$ and $P_m$ values, we identified the bifurcations responsible for magnetic field generation and reported a complex dependence of the dynamo on rotation. 
Regarding the question whether rotation favors or not the onset of dynamo (see, e.g., Meneguzzi and Pouquet \cite{meneguzzi1989turbulent} and Cattaneo and Hughes \cite{cattaneo2006dynamo}), for our ranges of control parameters it was clear from figure~\ref{Prandtl-magnetico-critico} that rotation has a strong impact on it, with the critical magnetic Prandtl number varying from 9.5 for $Ta=5$ to 5.5 for $Ta=1000$, but growing again for even larger $Ta$, reaching $P_m^c=7.5$ for $Ta=2500$. Thus, the critical $P_m$ depends nonlinearly on $Ta$. 

Our work confirms some universal results connected with RBC. For instance, the route to chaos via quasiperiodicity as a function of $R$, with the system subsequently moving back to periodicity for intermediate $R$ and, then, to chaos for higher $R$ was previously reported in detailed 2D \cite{paul12,oteski15} and 3D \cite{Podvigina2006magnetic} numerical simulations of RBC without rotation. On the other hand, our results on transition to dynamo through a nonhysteretic blowout bifurcation are in stark contrast to the hysteretic blowout bifurcation transition found in helically-forced turbulence in a periodic box, reported by Oliveira et al. \cite{oliveira2021chaotic}, where intermittency was not found. As a future work, it is of interest to study if similar behaviour is demonstrated by the rotating convective dynamos in the spherical geometry -- a configuration closer to astrophysical and geophysical problems.

The physical mechanisms responsible for magnetic field generation in planets and stars are extremely challenging for mathematical modelling -- even modern supercomputers are not powerful enough to resolve all spatial and temporal scales. Thus, simplified models must be considered to describe the mechanisms of magnetic field generations, e.g., plane layer dynamos. 
We stress that our model captures the main properties of  real astrophysical and geophysical MHD systems -- rotation, heat transfer and electromagnetics. However, in order to study more realistic systems, we plan to improve our models substituting the idealisations by more physically feasible assumptions. First, the stress-free~(\ref{eqn:6}) and perfect conductor~(\ref{eqn:7}) boundary conditions on the horizontal planes can be substituted by the no-slip insulating boundaries, as in \cite{tolmachev}. Second, we plan to vary the size of the convective cells (here only square convective cells with $L=4$ in (\ref{eqn:per}) were considered), because in the recent study~\cite{simon} of the same MHD system, it was found that smaller and rectangular cells were beneficial for magnetic field generation. The most beneficial size of the cell in the kinematic dynamo framework can be found solving the perturbation problem for Bloch eigenmodes~\cite{rjes1,rjes2}. We plan to continue our research on convective dynamos in these directions.

\ack
DNO acknowledges Aeronautics Institute of Technology – ITA.  RC acknowledges the financial support of the Foundation for Science and Technology (FCT/MCTES, Portugal) in the framework of the Associated Laboratory -- Advanced Production and Intelligent Systems (AL ARISE, ref. LA/P/0112/2020), the R\&D Unit SYSTEC (Base UIDB/00147/2020 and Programmatic UIDP/00147/2020 funds), and projects RELIABLE (ref. PTDC/EEI-AUT/3522/2020) and MLDLCOV (ref. DSAIPA/CS/0086/2020, financed by the program INCO.2030 -- National Initiative for Digital Competences e.2030). A part of the simulations was carried out with the OBLIVION Supercomputer (at the High Performance Computing Center, University of Évora) funded by the ENGAGE SKA Research Infrastructure (reference POCI-01-0145-FEDER-022217 - COMPETE 2020 and the FCT, Portugal) in the framework of the FCT call (3rd edition) for computational projects (ref. 2022.15706.CPCA.A2).
ELR acknowledges Brazilian funding agency CNPq, under grant 306920/2020-4. The work of FFF is funded by the CNPq Brazillian agency (project no. 401425/2023-1 - FAPEG/CNPQ Public Call No. 09/2022- PROFIX-JD). 

\section*{References}
\bibliography{references}

\end{document}